\documentclass[10pt,journal,compsoc]{IEEEtran}

\ifCLASSOPTIONcompsoc
  \usepackage[nocompress]{cite}
\else
  \usepackage{cite}
\fi

\ifCLASSINFOpdf
  \usepackage[pdftex]{graphicx}
\else
\fi

\begin{document}

\title{Simulating and visualizing COVID-19 contact tracing with Corona-Warn-App for increased understanding of its privacy-preserving design}

\author{Nikolas Gritsch,
        Benjamin Tegeler,
        and~Faheem Hassan Zunjani,~\IEEEmembership{Elite Masters in Data Science, Ludwig-Maximilians-Universit\"{a}t M\"{u}nchen}
}

\IEEEtitleabstractindextext{%
\begin{abstract}
The world is under an ongoing pandemic, COVID-19, of a scale last seen a century ago. Contact tracing is one of the most critical and highly effective tools for containing and breaking the chain of infections especially in the case of infectious respiratory diseases like COVID-19. Thanks to the technological progress in our times, we now have digital mobile applications like the Corona-Warn-App for digital contact tracing. However, due to the invasive nature of contact tracing, it is very important to preserve the privacy of the users. Privacy preservation is important for increasing trust in the app and subsequently enabling its widespread usage in a privacy-valuing population. In this paper, we present a visual simulation of the working of the Corona-Warn-App to demonstrate how the privacy of its users is preserved, how they're notified of infectious contacts and how it helps in containing the spread of COVID-19. 
\end{abstract}

}

\maketitle

\IEEEdisplaynontitleabstractindextext

\IEEEpeerreviewmaketitle

\IEEEraisesectionheading{\section{Introduction}\label{sec:introduction}}

\IEEEPARstart{T}{he} world has been grappling with the Coronavirus-Disease 2019, abbreviated as COVID-19, since December 2019. COVID-19 is a highly contagious disease caused by the Severe Acute Respiratory Syndrome Coronavirus 2 (SARS-CoV-2) and was first identified in Wuhan, China \cite{YUKI2020108427}. According to the World Health Organization (WHO), globally, as of 17 January, 2022 there have been 326.3 million confirmed cases of COVID-19 which resulted in 5.5 million deaths \cite{WHO_dashboard}. 

When dealing with a highly infectious respiratory disease of pandemic proportions, governments have several public health tools available to them. One of those tools which is critical and highly effective in containing and/or breaking the chain of infections is contact tracing \cite{contact_tracing}. 

\subsection{Digital Contact Tracing}
Contact tracing is a public health measure in which people who might have come into contact with an infected person are identified and information about them is collected. By testing, isolating and if needed, treating these contact persons, contact tracing aims to contain the spread of the infection. 
Contact tracing can be really effective in containing a communicable disease. For example, smallpox was eradicated not through universal immunization but by exhaustive contact tracing \cite{scutchfield2003principles}. Prior to COVID-19, contact tracing was largely performed manually with the help of public health officials.

Manual contact tracing involves the following steps:
\begin{enumerate}
  \item Identifying an infected person.
  \item Interviewing the infected person and recording their recent close contacts. Close contacts could also be further interviewed depending on the disease. 
  \item Contact persons are isolated, tested and provided care if found to be infected. 
  \item If found to be infected, the contact person is treated as an infected person and this process is repeated. 
\end{enumerate}

There are several potential drawbacks of manual contact tracing. The infected person might not remember their recent activities and contacts especially if it spans several days. They might also deliberately withhold information. Additionally, the contact persons might not be immediately reachable and in the meantime, they might carry the infection on to even more people. Furthermore, manual contact tracing is labour intensive and time consuming, requires a large financial investment, and is not an easily scalable technique. 

Today, a large part of the population in a number of countries carries a digital device like a smartphone in their pockets. Since a smartphone is usually always carried around by people, it can be a very good proxy for the person themselves when it comes to determining the whereabouts of that person. Digital contact tracing also does not suffer from some of the drawbacks of manual tracing like relying on the memory and honesty of the infected person. It is also possible to immediately notify contact persons when an infection is detected in one of their previous contacts. Risk notifications can therefore be efficiently propagated to the users. This would enable them to isolate themselves quickly, get tested and avoid spreading it to more people before they develop symptoms. Digital contact tracing is also scalable and thus can easily cover a very large population in a short amount of time.   

The concept of digital contact tracing has been around since 2007 \cite{digital_tracing_origin} achieving prominence in the past 2 years due to COVID-19. Not only has their effectiveness been challenged \cite{lancet_tracingapps}, due to the intrusive nature of tracking tools, they have also been subject to criticism raising ethical issues. One of the issues that has been a central point of recent discussion on contact tracing apps is user privacy.  

\subsection{Privacy in Digital Contact Tracing}
Due to recent large-scale data breaches which have lead to sensitive user data being vulnerable, there is a heightened consciousness about data privacy when using apps in general and tracking apps in particular. Different contact tracing apps collect different kinds of data and use different algorithms to process this data in order to generate risk notifications. The severity of privacy violation when using these apps depends, among other things, on the type and amount of data collected from its users and the location of processing of this data (client-side or server-side). 

Researchers have noted two primary approaches \cite{covid_digitalprivacy}:   
\begin{enumerate}
    \item \textbf{privacy-first} approach - protects user data by sacrificing data access to public health officials and researchers.
    \item \textbf{data-first} approach - collects data valuable to the authorities and researchers by intruding the users' privacy
\end{enumerate}

Germany's official contact tracing app, Corona-Warn-App, uses the privacy-first approach. The mechanisms used by the app for privacy preservation are explained in section \ref{corona_warn_app}. Due to its privacy preserving design, we have selected it as our application of choice for the purpose of this project.

\subsection{Visualizing the algorithm}
A lot of the arguments defending privacy-preserving tracing apps use highly technical descriptions which might not be immediately and easily understandable to a part of the target user base. The public perception of Corona-Warn-App remains unclear; although the German Twitter-sphere seems to largely support the adoption of the app \cite{cwa_public_perception}. As widespread adoption is critical to the success of contact tracing, it is important to develop trust and enable a general understanding of the inner workings of the privacy preserving mechanisms of these apps \cite{cwa_trust}. 

Humans are great at visually recognising patterns making visualizations a great tool for explaining complex concepts. Therefore, we have designed an easy-to-use simulation-based visualization which clearly demonstrates what data of the user and their contacts the Corona-Warn-App stores and processes, and how it notifies the users when an infection is recorded in their contacts. 

In the following sections, we begin by introducing the Corona-Warn-App, its functionalities and comparison to its peers in section \ref{corona_warn_app}. We also discuss the possible attacks that could lead to privacy violations. In section \ref{solution}, we introduce our simulation-based visualization and describe its implementation. Finally, in section \ref{conclusion}, we summarize our findings and potential future work in this project.

\section{Corona-Warn-App}
\label{corona_warn_app}

\subsection{International Overview}
There are a total of 49 different tracking apps in active use worldwide. Filtering by those which keep the identity of the users anonymous and have methods in place to delete recorded data once they are no longer useful and thereby follow basic privacy standards, leaves only 24 remaining. Out of those only 3 use GPS-Locations and all others are Bluetooth based. Five of them are centralized and all others are decentralized \cite{covid_tracing_tracker}.

The aforementioned privacy-first approach is strongly aided by a decentralised setting, as the majority of data remains on the user device and is not processed centrally. This raises the difficulty for large-scale data analysis.

In contrast, centralised systems require all users to transfer major parts of their data to an authority which then processes these - with the obvious privacy risks of movement relation data of possibly millions of people being collected and the resulting vulnerability to governmental misuse and data leakage due to implementation errors or hacking attacks. 
This was seen in the case of the Luca-App, which is a private app used in Germany in order to enable digital contract tracing at events or restaurants. Its data was accessed illegally by law enforcement in at least one case and network analysis indicates that it is possible to uniquely re-identify mobile devices and associate all their visited places to them - which cannot be verified as major parts of its source code are not open source \cite{luca_tracking,luca_police,luca_problems}.

There is a similar split in regards to the method with which the meeting of two users is determined. It is possible to do so using Bluetooth, which has very limited range and only allows for the knowledge that two users were within a certain distance at a certain time, thereby being privacy-first. 
In contrast actual location determination via either triangulation of cell-towers or using GPS are not only plagued by the accuracy of these \cite{gps_precision} but also by the actual location of a user being known, which allows for the easy building of movement profiles and thereby following a data-first approach \cite{covid_tracing_tracker}.

\subsection{The Corona Warn App}
After heavily criticised early plans of implementing a centralized tracing app, the German government decided on a decentralized Bluetooth approach, which is designed with the privacy-first goal \cite{covid_tracing_tracker}. The app itself, its framework and underlying platform were implemented by SAP and Deutsche Telekom is responsible for operating the back-end, and the network and mobile technology. It uses the Privacy-Preserving Contact Tracing specifications by Apple and Google (GAEN Framework \cite{apple_google_framework}), which ensures interoperability between the two major mobile operating systems and is itself designed privacy-first and completely open source in order to allow for outside scrutiny \cite{cwa_documentation}.

Each users phone acts as a Bluetooth beacon and regularly sends out a random identifier, Rolling Proximity Identifier (RPI), which is additionally switched out every 10-20 minutes in order to further increase privacy. 
These are derived via a hash function out of the Temporary Exposure Key (TEK), which is changed every 24 hours, and also uploaded in case of a positive test.
Every phone also listens for other users broadcasts and records and securely stores their RPI's. 
If a user realises that they were infected due to a test and decide to publish this in their CWA, their respective TEK's for the relevant time frame are uploaded to the back-end servers together with a Transmission Risk level (TRL), which is determined based on the symptom onset and thereby present infectiousness at the given time.
Every device retrieves all identifiers of infected users together with their TRL and compares them to their locally stored list of identifiers in order to compute the risk contained in their encounter. Every contact is bucketed into close, medium, far and very far buckets based on the strength of the Bluetooth signal and then the time spent at that strength multiplied with the weight for each attenuation bucket. Those values are then multiplied by the Transmission Tisk Value (TRV) which is assigned to each TRL and yields the normalised exposure times. All of these are then added together and their total sum is then bracketed into low risk, low risk with encounters and high risk.    
All thresholds in these computations are regularly pulled from the control servers, allowing for tuning of these values based on new scientific insights without the need for deploying a new version. 
An important factor is that border crossing travel is very common within the EU, so tracing apps should be able to interoperate in order for them to encompass these users \cite{cwa_SOLUTION_ARCHTIECTURE}.
There is a total of 22 contact tracing apps in the EU, only 2 of which (namely France (which also has a central system) and Hungary) are not technically able to interface with the others as they do not deploy the GAEN Framework. 2 of those which could (Portugal and Estonia) do not yet. Additionally, the Swiss tracing app is also connected to the Corona Warn App \cite{tracing_apps_europe, cwa_SOLUTION_ARCHTIECTURE}.

\subsection{Privacy evaluation and attack vectors}
The Chaos Computer Club, a well-known German Hacker and civil liberties group, has published 10 societal and technical requirements for contact tracing apps, which encompass privacy-by-design and must be fulfilled in their eyes to avoid "creating a privacy disaster". \cite{ccc_tracing_apps}

The technical aspects of these were evaluated by the developers of the CWA and the app was found in accordance.

\subsubsection{No Central Entity to Trust}
There is no secret information processed on the publicly accessible central servers and identifying metadata of the diagnosis keys are scrubbed before being processed in the back-end server. Any risk computation is carried out locally, with the actual instances of contact being stored locally within the GAEN framework and not accessible by the app itself \cite{CWA_CCC_COMPARISON}.

\subsubsection{Data Economy/Minimization}
Only necessary data is collected (most importantly location is and cannot be collected) and the diagnosis keys deleted after the COVID-19-relevant period of 14 days \cite{CWA_CCC_COMPARISON}.

\subsubsection{Anonymity}
Users are completely anonymous as long as their TEK's remain solely on their device. The RPI's recorded by other devices can only be mapped to a given contact if their TEK is known, which they have to willingly share. It is possible (and necessary) to attribute the corresponding RPI's of a day to a TEK once it has been published. However this TEK cannot be linked to a user or their device's IMEI without access to the secret storage of the device.
One possible attack vector could be: One could follow a given person and collect all their RPI's and thereby connect them to the person and then check whether the person is ever marked as infected. This would require a huge amount of effort in order to gain the additional information that they have been infected - as their movements are already known by following them \cite{CWA_CCC_COMPARISON}.

\subsubsection{No Creation of Central Movement or Contact Profiles}
The RPI's are never stored centrally and cannot be linked to one user, unless the TEK is uploaded where they are also only linked to a diagnosis key for a day and not the actual user.
TEK's are also only published once a certain threshold is reached, to mitigate the risk that there is only one person reporting positive for a prolonged period of time \cite{CWA_CCC_COMPARISON}.

\subsubsection{Unobservability of Communication}
The communication to the server is encrypted by HTTPS such that the message payload is not readable by outside observers and the metadata is removed before being processed in the servers, thereby making them not linkable on a database level. Static public key pinning ensures that communication only happens between the Corona-Warn-App and the server and reduces the possibility of man-in-the-middle attacks. 
Furthermore, fake messages are regularly sent in order to create an even message stream even in the case of someone monitoring a devices network traffic and deducing that contact to the server indicates a positive test \cite{CWA_CCC_COMPARISON}.

\section{Explaining and visualizing contact tracing through a simulation}
\label{solution}

\subsection{Idea}

To explain how the Corona-Warn-App works to a broad audience, text can be a sub-optimal medium. If it assumes too much prior knowledge, it can be overwhelming to readers, if it assumes too little, it can seem overly verbose. Therefore, we chose to create an interactive simulation\footnote{The code for our simulation is available at https://github.com/al-kindi/deds-corona-warn}, which can be more intuitive, insightful and engaging. Furthermore, the simulation can also be understood by people who are not familiar with the technical terms from the field of contact tracing.
The simulation should highlight three core points: \begin{enumerate}
    \item How the app works
    \item Why the app can prevent outbreaks
    \item How the app preserves the users' privacy
\end{enumerate}

\begin{figure*}[h]
\centering
\includegraphics[width=18cm]{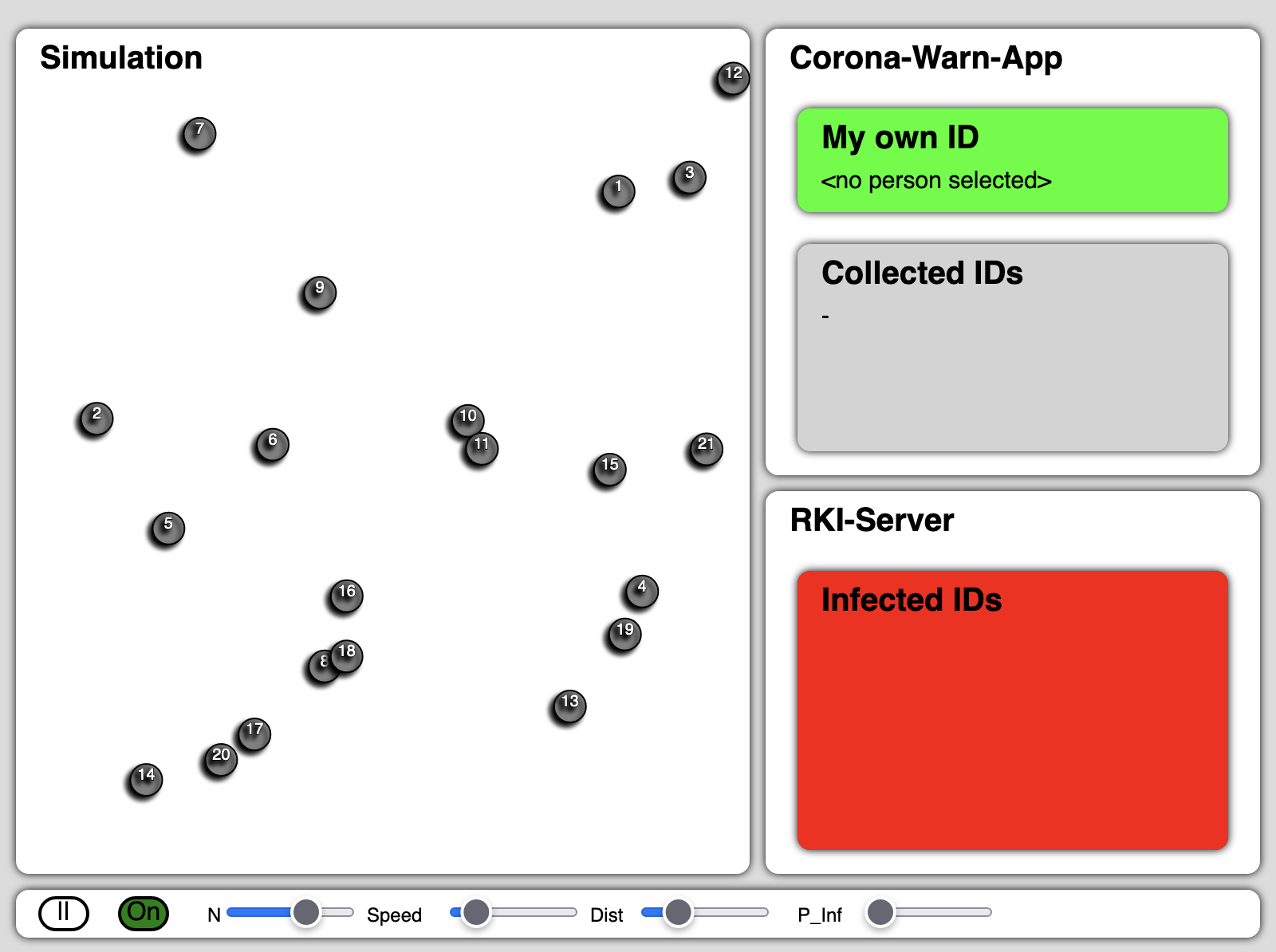}
\caption{Overview of the simulation of the Corona-Warn-App}
\label{fig_simulation_overview}
\end{figure*}

\subsection{Implementation}
We decided to write the simulation in JavaScript so that it can be easily hosted as a website. Furthermore, we decided to use the p5.js library, as it provides convenient wrappers for visual elements in JavaScript. To run the simulation, the HTML file can be either opened in a browser or served via a web server.

Figure \ref{fig_simulation_overview} shows the simulation shortly after starting it. It is divided in three sections: on the left, the actual simulation of the population can be seen. On the top right, the contents of the Corona-Warn-App can be seen. By clicking on an individual, the data that the app of this individual stores are displayed. On the bottom right, the information that the Robert-Koch-Institute (RKI) server stores is displayed.

\begin{figure}[h]
\centering
\includegraphics[width=8cm]{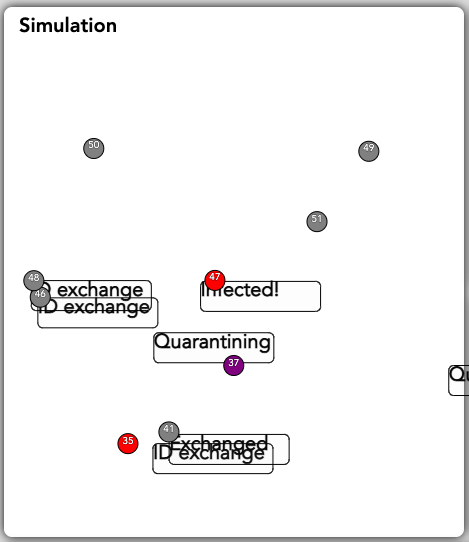}
\caption{Healthy (grey), infected (red) and quarantining (purple) individuals in the simulation}
\label{fig_simulation_part}
\end{figure}

Each individual in the population simulation is represented by a circle. The color of the circle represents the health status of the individual, Figure \ref{fig_simulation_part} shows multiple individuals with different health statuses.
Initially, each circle is grey representing a healthy individual. If an individual gets infected with COVID-19, its circle first turns red. During this period, the person can infect other nearby people but does not know yet that it is infected. After a certain time, the individual will develop symptoms and therefore take note of its infection. This is indicated by a color change from red to purple. If this happens, the person will immediately go to quarantine. This is visualized by the individual leaving the screen to the left. If the individual is not quarantining because it has developed symptoms, but because it got a warning from the app and confirmed this with a positive test, it does not change its color to purple and it will leave the screen to the right side instead of the left side. If a circle has left the screen, it is in quarantine and will not pose a further danger to others, therefore it is removed from the simulation. Quarantining individuals that left the simulation will be replaced by new, healthy ones, if there are less than 60\% of the original population left.

\begin{figure}[h]
\centering
\includegraphics[width=8cm]{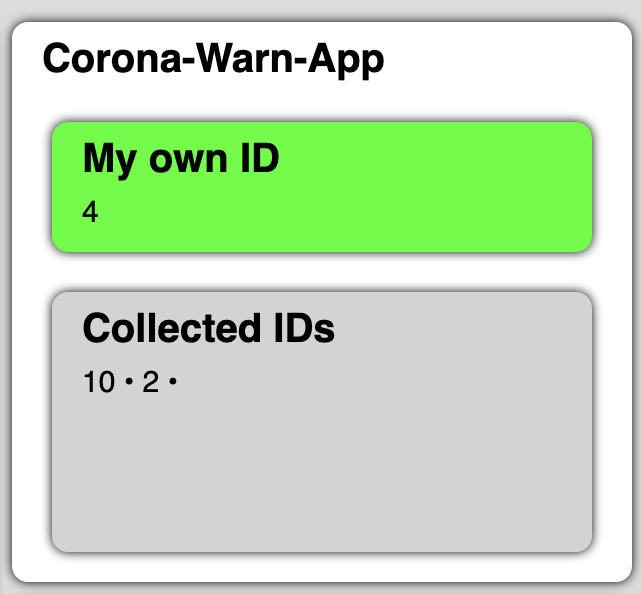}
\caption{The Corona-Warn-App in the simulation}
\label{fig_cwa_part}
\end{figure}

The individuals move around randomly. Each person in the population can get infected randomly with the virus with a small chance and become "patient zero". After that, they can infect people that come closer than a certain radius. Whenever two people in the simulation get at least that close, they also exchange a signal in the Corona-Warn-App: each person sends their unique ID to the other person, which stores it in a list of collected IDs. The display of the Corona-Warn-App is shown in Figure \ref{fig_cwa_part}. Different individuals can be selected by clicking on them.

\begin{figure}[h]
\centering
\includegraphics[width=8cm]{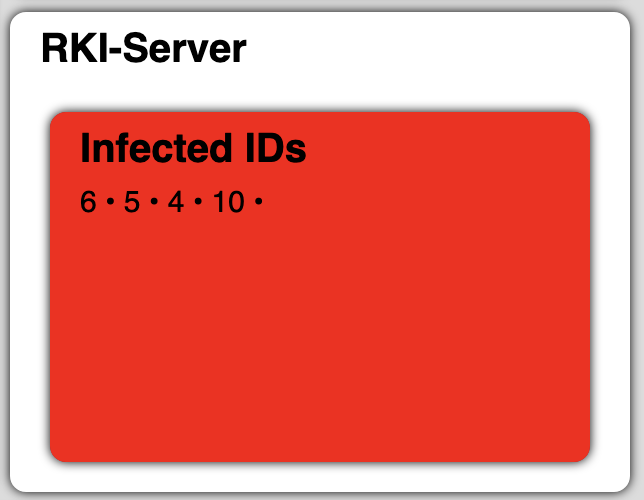}
\caption{The RKI server display with some IDs of infected individuals}
\label{fig_rki_part}
\end{figure}

If a person develops symptoms, they immediately notify the RKI server of their infection by publishing their own ID to the server. The server display is shown in Figure \ref{fig_rki_part}. All individuals will regularly poll the server for the list of infected IDs. If the see an ID on the server that they have in their private list of collected IDs, they will take a test and, if they receive a positive result, immediately quarantine themselves by moving out of the screen to the right.

\begin{figure}[h]
\centering
\includegraphics[width=10.5cm]{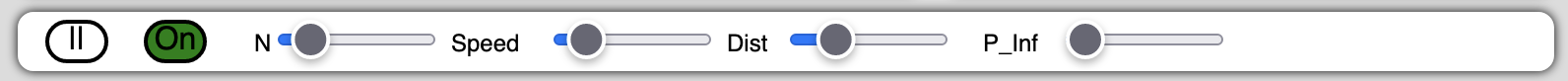}
\caption{Control panel for simulation parameters}
\label{fig_params_part}
\end{figure}

Several parameters of the simulation can be adjusted interactively. The control panel is shown in Figure \ref{fig_params_part}. The simulation can be stopped and started again with the first button. The second button controls if the population uses the app or not. The following four sliders control the number of individuals in the population (with more individuals leading to spaces being more crowded and therefore a higher risk of infection), the speed at which the individuals move (with higher speed leading to more interactions and therefore a higher risk of infection), the distance at which infected individuals can spread the virus (the higher the distance, the more infectious the individuals will be) and the natural rate of outbreak of the virus within the population (the higher this rate is, the more different clusters will exist).

\subsection{Necessary simplifications}

The natural infections represent infections from outside of the simulated population, not actual natural infections via mutation or transmission from animals. Also, individuals do not randomly take tests, but only when the app prompts them to do so. Moreover, the distance at which individuals can infect others is assumed to be the same as the distance in which the Corona-Warn-App will warn others. In the real world, this would be optimal, too, but is probably not always fulfilled. Also, in the simulation of the app, infections do not work gradually but instantaneously, and there is no differentiation by length of exposure, while in the real app, a certain number of consecutive signals by the infected individual must have been received for the contact to be considered a high risk contact.

The most important simplification of the simulation concerns the privacy concerning aspects of the Corona-Warn-App. In the real app, the IDs exchanged by the devices are randomly generated, change frequently and do not allow personal identification of a user. This proved to be too much complexity for our simulation, bloating the UI and obfuscating the tracing mechanism while also not allowing users to see in an easy way where the collected IDs come from. Therefore, we decided to simplify this feature and give our simulated individuals IDs that are permanent and therefore theoretically allow personal identification. We think however, that the overall explanatory ability of our simulation increased through this decision.

\subsection{Insights}

Our simulation shows very well that the whole population gets infected very fast after an initial outbreak if the Corona-Warn-App is not used. If the app is used, the simulation shows well that clusters of a new outbreak grow at first, while the patient zero has not noticed their infection but already infected others. However, when the first individual in a cluster develops symptoms, the whole cluster is sent to quarantine very fast, stopping the further spread of the disease. Another interesting aspect is that for high values of the number of individuals, the speed, or the infection radius, nearly the whole population will get infected despite using the app. This proves that while the app is helpful in reducing the spread of the virus, it is still important that people avoid crowded places, reduce their overall contacts to a reasonable amount and stick to hygienic measures like wearing masks (which reduce the radius of infection).

\section{Conclusion and Future Work}
\label{conclusion}
By creating a population simulation of individuals using the Corona-Warn-App, we show intuitively how the app can effectively contain outbreaks and send clusters of infected individuals to quarantine without keeping a central record of individuals' interactions. This especially shows that privacy-first approaches are on-par with data-first approaches. By making the simulation interactive, we allow viewers to get a deeper understanding of the population dynamics of the spread of COVID-19 - with and without the app. Analyzing the effect of different parameters on the dynamics of the simulation, we conclude that in addition to using the app, social distancing measures might be necessary as well.

In the future, the simulation could be improved in several ways. Improving the user-interface of the simulation might make the interface look cleaner and the pop-ups more readable. Scenario-based simulation templates can be programmed which would allow the users to simulate real-world settings such as a supermarket, classroom, subway ride, long-distance train, restaurant, etc. Additionally, staggering depths of the actual functional complexity could be revealed, allowing for deeper insights into the actual technical frameworks without having to read all of the documentation, which is not always easily understandable to non-technicians. Overall, we believe that our visualization is an important first-step in enabling the general population to better understand the functioning of digital contact tracing apps like the Corona-Warn-App. We hope that such a simulation increases trust in the software by showing its privacy preservation, leading to its widespread adoption and eventual success as a public health measure in breaking the chain of infections. 


\ifCLASSOPTIONcompsoc
  \section*{Acknowledgments}
\else
  \section*{Acknowledgment}
\fi
The authors would like to thank Prof. Dr. Dieter Kranzlm\"{u}ller, Mr. Jan Schmidt and Mr. Fabio Genz for giving them the opportunity to work on the topic of "Breaking the chains of infection - COVID-19 contact-tracing apps" as part of the Summer Semester 2021 course "Data Ethics and Data Security".

\ifCLASSOPTIONcaptionsoff
  \newpage
\fi

\bibliographystyle{IEEEtran}
\bibliography{refs}



\end{document}